\documentclass{article}

\usepackage{arxiv}

\usepackage[utf8]{inputenc} % allow utf-8 input
\usepackage{hyperref}       % hyperlinks
\usepackage{url}            % simple URL typesetting
\usepackage{booktabs}       % professional-quality tables
\usepackage{amsfonts}       % blackboard math symbols
\usepackage{nicefrac}       % compact symbols for 1/2, etc.
\usepackage{microtype}      % microtypography
\usepackage{lipsum}		% Can be removed after putting your text content
\usepackage{graphicx}
\usepackage{natbib}
\usepackage{doi}
\usepackage{newtxtext}
\usepackage{newtxmath}

\title{Experimental determination of the sodium K-shell atomic fundamental parameters for X-ray spectroscopy}

\author{\href{https://orcid.org/0009-0000-4251-3283}{\includegraphics[scale=0.06]{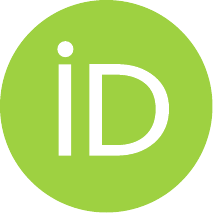}\hspace{1mm}Marie-Louis Venzke}\\
   Helmholtz-Zentrum Berlin (HZB)\\
   Hahn-Meitner-Platz 1\\
   14109 Berlin, Germany\\
\And 
\href{https://orcid.org/0000-0002-4446-5518}{\includegraphics[scale=0.06]{orcid.pdf}\hspace{1mm}Katja Frenzel} \\
Physikalisch-Technische Bundesanstalt (PTB)\\
	Abbestr. 2-12 \\
	10587 Berlin, Germany\\ 
\And 
\href{https://orcid.org/0000-0002-0712-903X}{\includegraphics[scale=0.06]{orcid.pdf}\hspace{1mm}Philipp Hönicke$^{1,2}$}\\
	1 : Helmholtz-Zentrum Berlin (HZB)\\
   Hahn-Meitner-Platz 1\\
   14109 Berlin, Germany\\
  \\
  2: Physikalisch-Technische Bundesanstalt (PTB)\\
	Abbestr. 2-12 \\
	10587 Berlin, Germany\\ 
\texttt{philipp.hoenicke@helmholtz-berlin.de} \\
}

\begin{document}
\maketitle

\begin{abstract}
Reliable fundamental parameters are essential for accurate X-ray fluorescence analysis of sodium. This work presents updated experimental values with reliable uncertainties for sodium K-shell fundamental parameters such as: fluorescence yield, photoionization cross sections, and Auger yields. Using a physically calibrated setup and a thin NaCl layer on a silicon nitride membrane, a holistic determination approach was applied to reduce uncertainties. The new values represent a significant improvement on the ones widely used in databases. All data are available via Zenodo to support precise sodium quantification in scientific and industrial applications. 
\end{abstract}


\section{Introduction}
Sodium, which is amongst the most abundant elements on earth \cite{Yaroshevsky_2006}, plays a major role in many areas of research, including life sciences \cite{Jaques_2023}, astronomy \cite{Shen_2023} and many others \cite{M_ller_2011}. 
Among the various applications, sodium-ion batteries (SIBs) \cite{Usiskin_2021} have gained particular attention as a promising alternative to lithium-based systems, especially for stationary energy storage. Their appeal lies in the high natural abundance and geographic accessibility of sodium, which translate to lower raw material costs and enhanced sustainability \cite{Liu2023,WU2024}. As interest in SIBs grows, so does the need for reliable and precise analytical methods to support their development. In particular, improving the accuracy of sodium quantification is essential for understanding interfacial processes, degradation mechanisms, and overall performance. Reducing uncertainties in sodium detection thus contributes directly to advancing SIB technology and supports the broader shift toward sustainable energy systems.

Accurate knowledge of key atomic fundamental parameters (FPs), such as fluorescence yields and photoionization cross sections, is essential for accurate quantitative x-ray fluorescence analysis (XRF) in various applications. The quantification results strongly depend on the precision of the used FP data and therefore, the availability of established FP values is of particular importance. Unfortunately, much of the available literature sodium FP data is outdated, only derived from interpolations of neighboring elements, or based solely on theoretical calculations without experimental validation. In addition, the uncertainties of these FP values are often unknown or estimated \cite{Krause1979}. To overcome this issue, there are initiatives such as the international Fundamental X-ray Parameters Initiative (FPI) \cite{FPI}, which is revising and improving the FP databases by means of new experiments \cite{Menesguen2017, Guerra2018} and new advanced calculations \cite{Baptista_2024}.
At Physikalisch Technische Bundesanstalt, calibrated instrumentation \cite{Beckhoff_2022} is used to perform experiments to either evaluate existing FP values or to redetermine them \cite{P.Hoenicke2016, P.Hoenicke2016a, Kayser_2022}. Recently, in order to further reduce the experimental uncertainties and to expand the set of accessible FPs, an updated holistic experimental and data evaluation procedure for improved experimental FP determination \cite{H_nicke_2023} has been developed. As shown in our previous works \cite{H_nicke_2023, H_nicke_2024, Wauschkuhn_2024}, the holistic approach provides a more reliable insight into the accessible FPs as compared to the earlier applied strategies. This work focuses on the determination of various sodium FPs, including the fluorescence yield and the photoionization cross sections of the sodium K-shell. All results are also available for download as plain text via Zenodo \cite{Hoenicke2025Zen}.

\section{Experimental section}

To experimentally determine the sodium K-shell FPs, such as the K-shell fluorescence yield $\omega_{\text{F}}$ or the K-shell fluorescence production cross section (FPCS) $\sigma_{\text{K}}(E_0)$, either a free-standing thin foil or a thin and homogeneous coating of pure sodium on a thin carrier is required. As sodium is highly reactive, both of these options cannot be realized. Thus, we are using a sodium containing chemical compound, namely sodium chloride (NaCl), coated onto a thin silicon nitride (SiN) membrane as a sample.  We have obtained a NaCl coating of nominally 102.5 $\mathrm{\mu g cm^{-2}}$ on a nominally 1000 nm thick SiN membrane from Micromatter Technologies Inc.\texttrademark.

Using this coated membrane, and an uncoated SiN membrane of identical thickness, we have performed transmission and X-ray fluorescence experiments for the incident photon energy range from below the sodium K-shell up to the silicon K-shell. These experiments have been carried out using an in-house developed ultra-high vacuum chamber \cite{J.Lubeck2013}, which was placed in the focal position of PTBs plane grating monochromator (PGM) beamline at BESSYII \cite{F.Senf1998}.

The UHV chamber was equipped with a calibrated window-less silicon drift detector (Bruker X-Flash), which was radiometrically calibrated in order to determine its detection efficiency \cite{F.Scholze2009} as well as the detector response functions for relevant photon energies. Furthermore, a calibrated apperture at a well-known distance to the sample is used to precisely define the solid angle of detection and calibrated photo diodes are used to determine and monitor the incident photon flux at each probed monochromatic photon energy.

\begin{figure}[!h]
	\centering
	\includegraphics[height = 6.5cm]{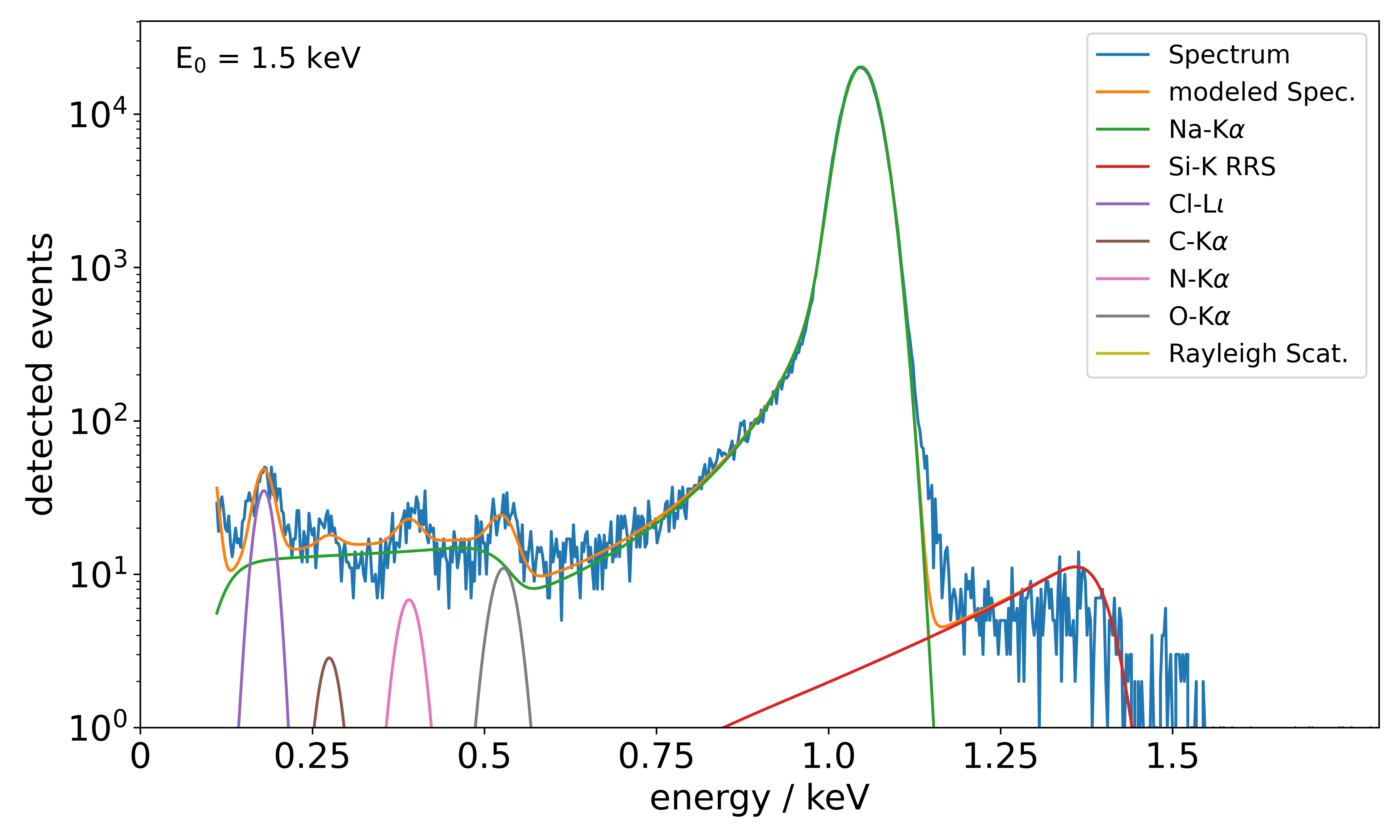}
    \caption{Example X-ray fluorescence spectrum of the NaCl coated 1000 nm thick SiN membrane recorded at an incident X-ray photon energy of 1.5 keV with the respective major deconvolution contributions.}
	\label{fig:fig0}
\end{figure}

The samples were aligned to be in the center of the experimental chamber by an x-y-scanning stage. For all experiments, both the incident and the detection angle were set to 45\textdegree. For the transmission experiments, the incident photon energy was varied in small steps (fractions of one eV) in the vicinity of the Na-K attenuation edge and in larger steps further away from this edge. For each incident photon energy, several readings of a photo diode placed in the beam behind the sample were recorded and averaged. The X-ray fluorescence experiments were performed only at incident photon energies above the Na-K edge and with less dense energy steps. Each X-ray fluorescence spectrum was recorded and later deconvolved employing the detector response functions for relevant fluorescence lines and relevant spectral background contributions. A comparison between the recorded spectrum and the deconvolution is shown in Figure \ref{fig:fig0} for an excitation photon energy of 1.5 keV.

\begin{figure}
	\centering
	\includegraphics[height = 6.5cm]{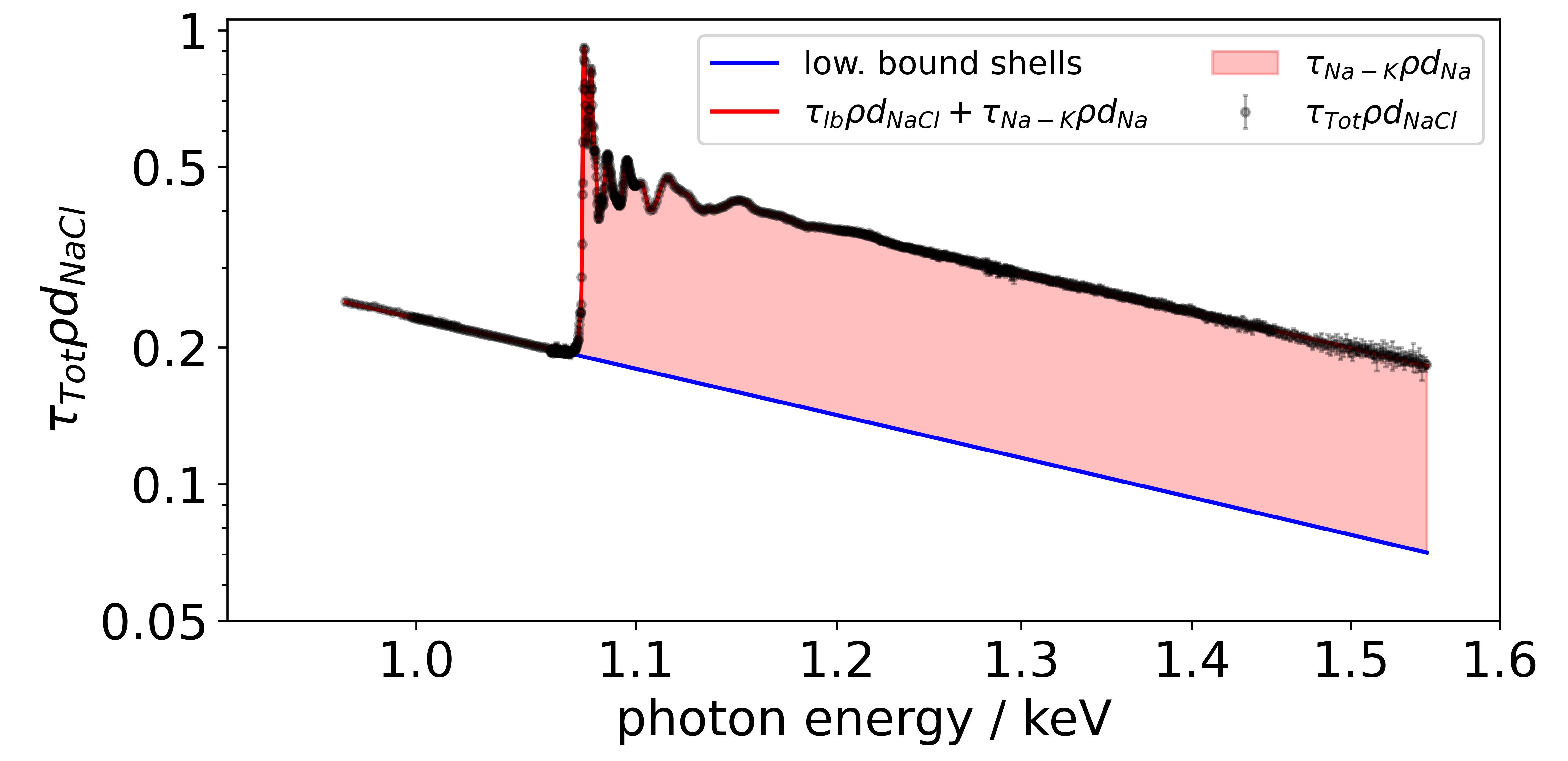}
    \caption{The total sample specific photoionization cross section $\tau_{Tot,NaCl}(E_0)\rho d$ derived from the transmission measurements and the isolation of the Na-K shell contribution is shown (see main text for further details).}
	\label{fig:fig0a}
\end{figure}

From the transmission data, the sample specific mass attenuation coefficients (product of the mass attenuation coefficient $\mu$, material density $\rho$ and thickness $d$) can be derived employing the Beer-Lambert law. In the case of the NaCl on SiN, it represents the sum of the sample specific mass attenuation of NaCl and SiN, respectively. Using the transmission data from the blank SiN membrane, the contribution for NaCl can then be isolated and the sample specific mass attenuation factors $\mu_{S}(E_0)\rho d_{NaCl}$ for NaCl can be obtained. As sodium compound coating was used, the mass attenuation coefficients of sodium cannot be directly derived from the dataset. Employing eq. \ref{eq:scatremoval} and X-raylib data for coherent, incoherent scattering and the mass attenuation of NaCl, the relative scattering contributions can be removed in order to derive the sample specific total photoionization cross sections $\tau_{Tot}(E_0)\rho d_{NaCl}$ for NaCl. In Figure \ref{fig:fig0a}, the thereby obtained total photoionization cross sections are shown including the isolation of the contribution of the Na-K shell ($\tau_{Na-K}(E_0)\rho d_{Na}$). This separation into the relevant partial subshell contributions is performed by scaling Ebel polynomials \cite{H.Ebel2003} into the $\tau_{Tot}(E_0)\rho d_{NaCl}$ dataset.

\begin{equation}
\tau_{Tot,NaCl}(E_0)\rho d = \mu_{S,NaCl}(E_0)\rho d - \frac{\sigma^{XLib}_{C,NaCl}(E_0) + \sigma^{XLib}_{I,NaCl}(E_0)}{\mu^{XLib}_{NaCl}(E_0)}
\label{eq:scatremoval}
\end{equation}

From these experimental results, the sodium K-shell fluorescence yield $\omega_{\text{F}}$ and subsequently also the sodium K-shell Auger yield $\omega_{\text{A}}$, as well as the Na-K subshell photoionization cross sections $\tau_{\text{K}}(E_0)$ and the FPCSs $\sigma_{\text{K}}(E_0)$ can be determined.  

To determine these FPs, we employ the K-shell adopted version \cite{H_nicke_2024} of the recently introduced holistic FP determination approach \cite{H_nicke_2023}. Unlike conventional FP data evaluation methods, this approach utilizes a significantly expanded photon energy range and an innovative combined data analysis scheme. As a result, more subshell-specific FPs are accessible and the achievable uncertainties can be reduced \cite{H_nicke_2023, H_nicke_2024, Wauschkuhn_2024}.

Using the Sherman equation \cite{Sherman1955} and solving it for the sample specific K-shell FPCS $\sigma_{\text{K}}(E_0)\rho d_{Na}$, we can gain access to $\omega_{\text{F}}$ (see eq. \ref{eq:prodCS}).

\begin{equation}
\sigma_{\text{K}}(E_0)\rho d_{Na} = \omega_{\text{F}} \tau_{\text{K}}(E_0)\rho d_{Na} = \frac{\Phi^d_{Na-K}(E_0)M_{Na-K}(E_0)}{\Phi_0(E_0)\frac{\Omega}{4\pi}}
\label{eq:prodCS}
\end{equation}
with
\begin{equation}
M_{Na-K}(E_0) = \frac{(\frac{\mu_S(E_0)\rho d}{\sin \theta_{in}}+\frac{\mu_S(E_f)\rho d}{\sin \theta_{out}})}{(1-\exp[-(\frac{\mu_S(E_0)\rho d}{\sin \theta_{in}}+\frac{\mu_S(E_f)\rho d}{\sin \theta_{out}})])},
\label{eq:M}
\end{equation}

Here, $\theta_{in}$ and $\theta_{out}$ are respectively the incident angle of the excitation photon beam and the detection angle of emitted the x-ray fluorescence radiation, $\Phi_0(E_0)$ is the measured incident photon flux, $M_{Na-K}(E_0)$ is the sample specific attenuation correction factor for the Na-K shell fluorescence radiation, $\frac{\Omega}{4\pi}$ the detection solid angle and $\Phi^d_{Na-K}(E_0)$ the measured fluorescence photon flux of Na-K shell fluorescence radiation. To determine the fluorescence photon flux $\Phi^d_{Na-K}(E_0)$, the deconvoluted detected events for the sodium K-shell fluorescence line $F(E_f)$ must be normalized by the integration time of each spectrum $t_{int}$ and the SDD's detection efficiency $\epsilon(E_f)$ at the corresponding photon energy of each fluorescence line. To determine $\omega_{\text{F}}$, knowledge of the sample areal mass (defined as the product of the density $\rho$ and the thickness $d$ of the NaCl layer) and on the exact stoichiometry is not required. Instead, only the sample-specific parameters given by the products of $\tau_{\text{Na-K}}(E_0)\rho d_{Na}$ and $\mu_S(E_0)\rho d_{NaCl}$ are needed and can be extracted directly from the recorded sample transmission data as shown earlier. The incident photon flux $\Phi_0(E_0)$ and the detection solid angle $\frac{\Omega}{4\pi}$ are determined either through measurements with calibrated photodiodes or by utilizing calibrated apertures integrated into the instrumentation \cite{Beckhoff2008}.\newline

The sample specific attenuation correction factor $M_{Na-K}(E_0)$ for both incident ($E_0$) and fluorescence radiation ($E_f$) is calculated using equation \ref{eq:M}, where the sample-specific attenuation coefficients $\mu_S(E_0)\rho d$ and $\mu_S(E_f) \rho d$ are determined by the transmission experiments, making $M_{Na-K}(E_0)$ independent of any mass attenuation coefficient database values.

Analogous to the holistic approach for L-subshell parameters \cite{H_nicke_2023, HfPaper}, the analysis utilizes an extensive transmission and fluorescence dataset with excitation energies well above the sodium K-shell. The holistic methodology involves an optimization procedure that simultaneously decomposes the sample-specific photoionization cross section into its constituent components and calculates $\omega_{\text{F}}$ using equation \ref{eq:prodCS} at each probed photon energy above the ionization threshold. The mean value of these derived fluorescence yields is then used to solve the same equation for the sodium K-shell $\tau_{\text{K}}(E_0)\rho d$. The optimization leverages the energy independence of the K-shell fluorescence yield, the consistency between transmission and fluorescence-derived $\tau_{\text{K}}(E_0)\rho d$, and the accurate representation of the measured total $\tau_{Tot,NaCl}\rho d$ by the sum of scaled subshell components. A Markov-Chain-Monte-Carlo (MCMC) algorithm \cite{Foreman_Mackey_2013} is employed, with the $A_i$ coefficients of the Ebel polynomial \cite{H.Ebel2003} representation of the photon energy dependent subshell photoionization cross sections as variable parameters. The optimization requires two scaling parameters for the lower bound shells (lb) of NaCl. For the sodium K-shell contribution, the number of variable parameters ($A_0$ to $A_5$) can range from one to six. Varying more parameters allows for adjustments to the photon energy dependence of the sodium K-shell contribution, while varying only $A_0$ modifies its magnitude. Fixed parameters of the Ebel polynomial retain their values from Ebel's original dataset for the sodium K-shell.

\section{Results and Discussion}

\subsection{Determination of the K-shell fluorescence - and Auger yields}
\begin{figure}[!h]
	\centering
	\includegraphics[height = 7cm]{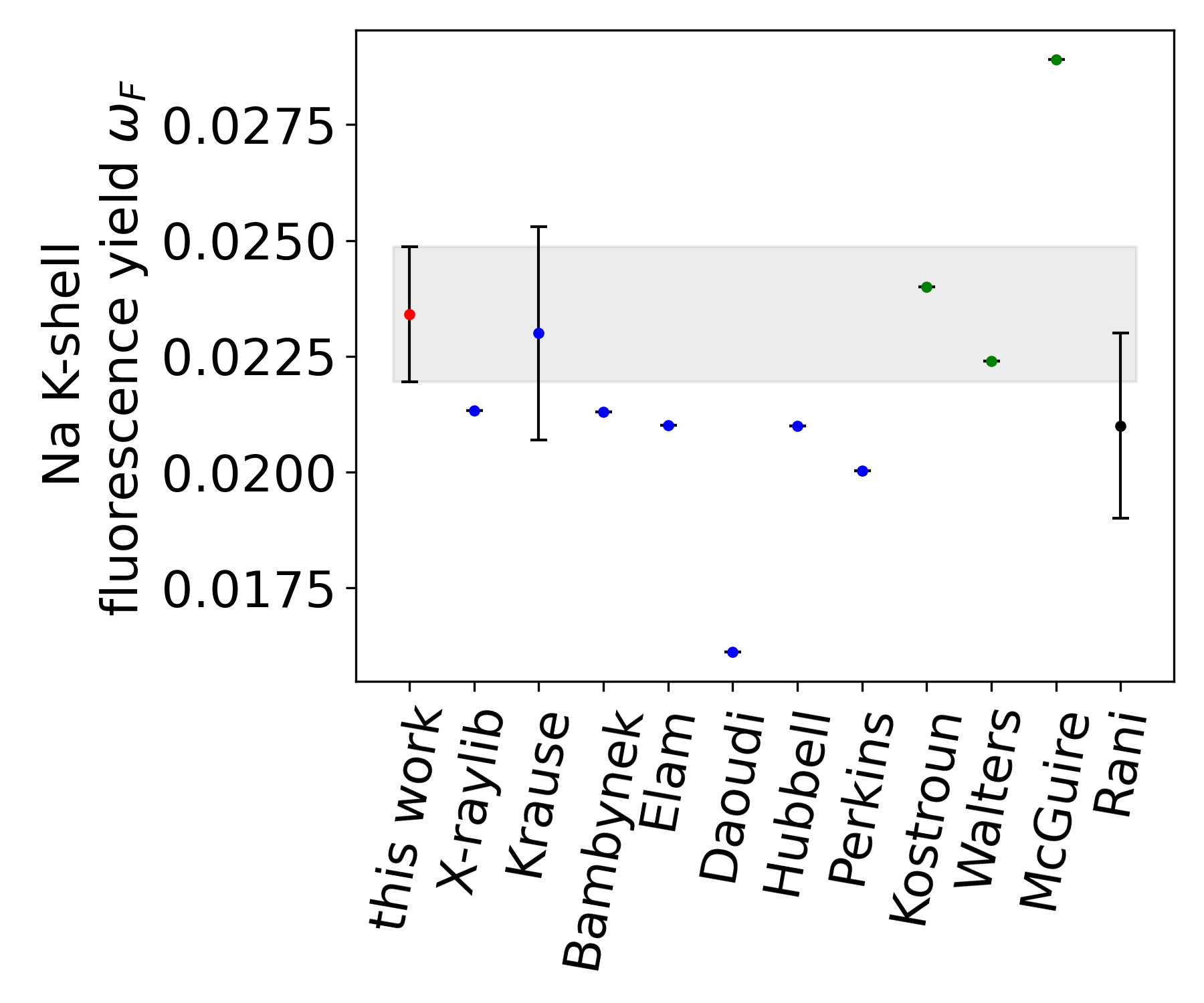}
    \caption{Comparison of the experimentally determined Na-K shell fluorescence yield with available data in the literature. This includes data from compilations (blue dots) such as X-raylib\cite{T.Schoonjans2011}, by Krause\cite{Krause1979}, by Bambynek\cite{Bambynek1984}, Elam\cite{W.T.Elam2002}, Daoudi\cite{S.Daoudi2015}, Hubbell\cite{J.H.Hubbell1994} and Perkins\cite{S.T.Perkins1991}, theoretically calculated data (green dots) by Kostroun\cite{V.O.Kostroun1971}, Walters\cite{Walters_1971} and McGuire\cite{McGuire_1969} and an experimentally determined value (black dot) by Rani\cite{A.Rani1988}.}
	\label{fig:fig2}
\end{figure}

In Figure \ref{fig:fig2}, the experimentally determined $\omega_{\text{F}}$ is shown in comparison to available data from various literature sources.  This includes the most common database collections, such as X-raylib \cite{T.Schoonjans2011}, Krause \cite{Krause1979}, Bambynek \cite{Bambynek1984}, Elam \cite{W.T.Elam2002} and others as well as theoretically calculated data and one available experimentally determined fluorescence yield value by Rani \cite{A.Rani1988}. For easier comparison, the uncertainty regime of our result is indicated as grey shaded box. Here, a value of $0.0234 \pm 0.0015$ was determined.

An agreement within the stated uncertainties (if available) is found for the data provided by Krause, the theoretical calculated values by Kostroun and Whalters and for the experimental result by Rani. Most of the other values are slightly lower. The interpolated value by Daoudi seems to be significantly too low and the calculated value by McGuire too large.

\begin{figure}[!h]
	\centering
	\includegraphics[width = 8cm]{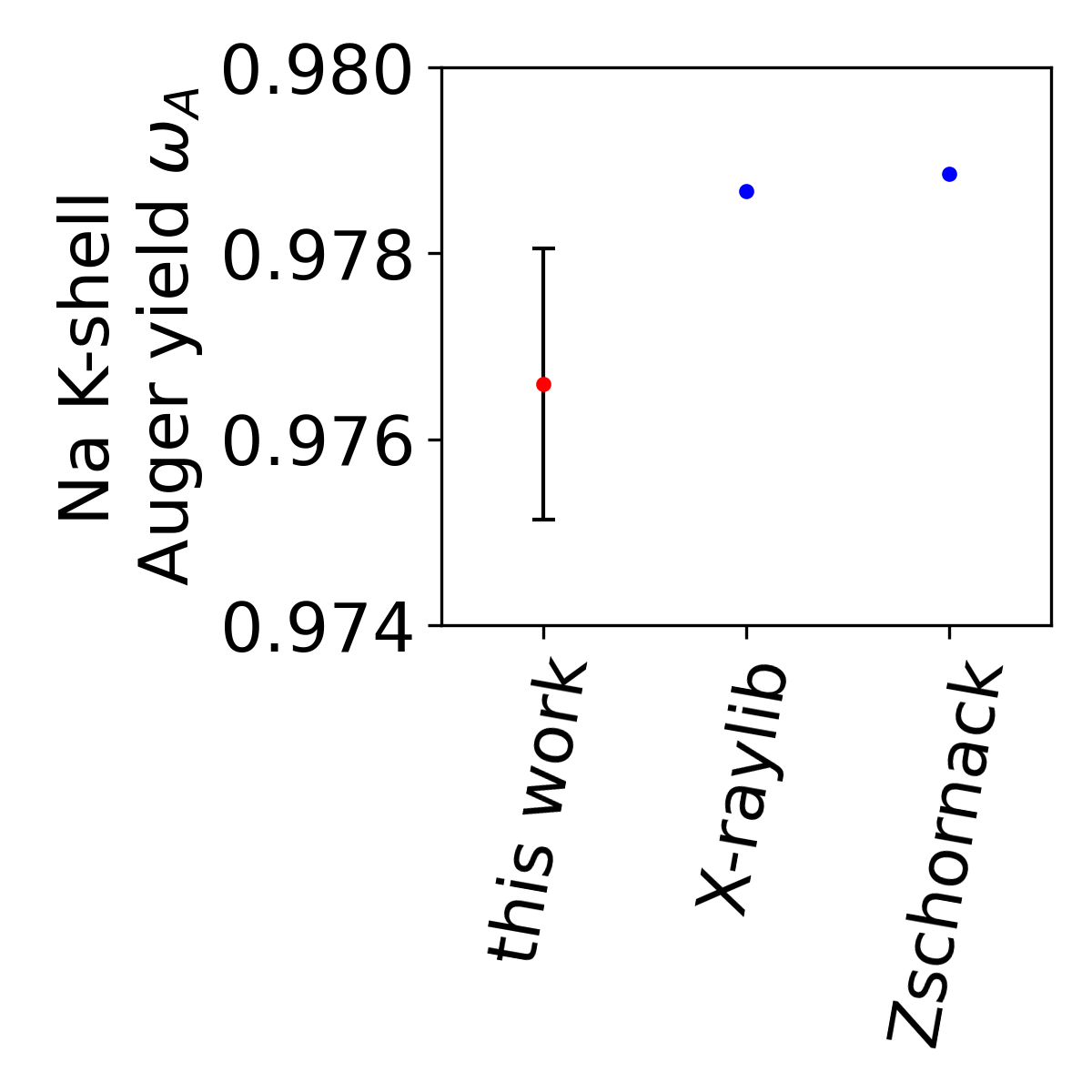}
    \caption{Comparison of the experimentally determined Na-K shell Auger yield with available data in literature compilations by Schoonjans (X-raylib)\cite{T.Schoonjans2011}, and Zschornack\cite{Zschornack2007}.}
	\label{fig:fig3}
\end{figure}

As the K-shell Auger yield and the K-shell fluorescence yield add up to be unity, the Auger yield can be calculated by subtracting the determined fluorescence yield result from one. The thereby derived sodium K-shell Auger yield is $0.9766 \pm 0.0015$ and is shown in Figure \ref{fig:fig3}. Consequently it shows opposite behavior as compared to the fluorescence yield value with the X-raylib result being slightly to large now. The result taken from Zschornack \cite{Zschornack} is of similar agreement to our result being slightly higher than the X-raylib value. But this may be affected by the fact that the Zschornack-result is based on a manual digitization of a plot.

\subsection{Determination of the K-shell fluorescence production cross sections}
\begin{figure}[!h]
	\centering
	\includegraphics[width = 12cm]{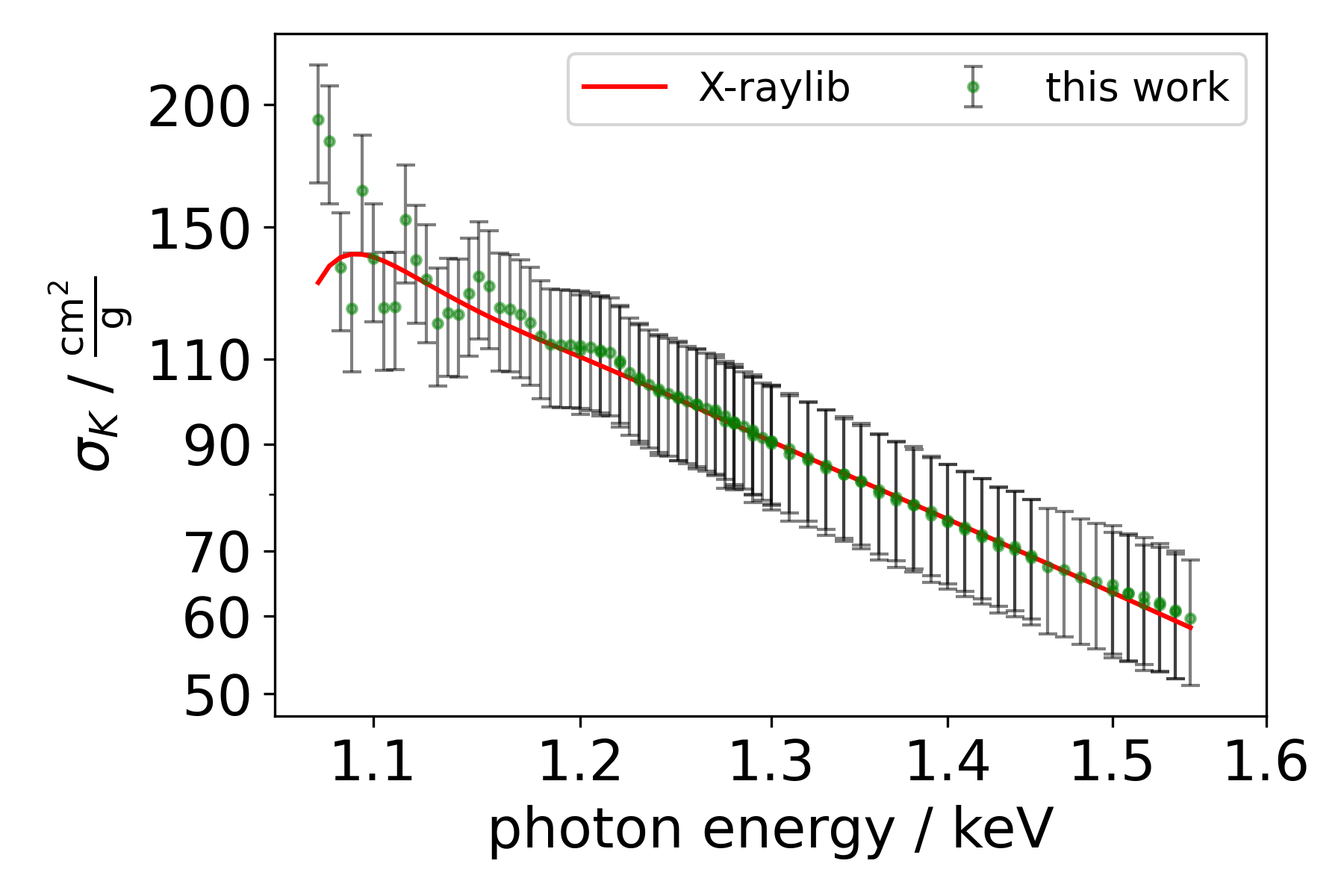}
    \caption{Comparison of the experimentally determined Na-K shell FPCSs as a function of the photon energy in comparison to data from X-raylib \cite{T.Schoonjans2011}.}
	\label{fig:fig4}
\end{figure}

The sample specific sodium K-shell FPCSs can be calculated using equation \ref{eq:prodCS}. By determining the sample's areal mass of sodium, employing a reference-free quantification at an incident photon energy far above any fine structure oscillations and tabulated FP data \cite{T.Schoonjans2011}, the absolute FPCS for Na-K fluorescence can be derived. The results are shown in comparison to X-raylib data in Figure \ref{fig:fig4}. This quantification of the samples areal mass of sodium results in a relatively large overall uncertainty of the derived FPCSs as well as matching absolute values with respect to X-raylib by definition. However, we have cross checked the determined areal mass of sodium by also deriving a result from scaling the tabulated mass attenuation coefficients from X-raylib\cite{T.Schoonjans2011} for Na and Cl to the derived sample specific mass attenuation coefficients. This yields a very similar result. The procedure can be further improved by employing an independent approach for the areal mass determination based on precision weighing and area determination in the future.

As the X-raylib dataset is for atomic sodium, the x-ray absorption fine structure in the vicinity of the sodium K-edge is missing in the data. As this is specific for the chemical state of the probed element within the present chemical compound, it would be rather complex to be integrated into such databases. However, the photon energy dependence of the experimental data towards higher photon energies is in good agreement to the X-raylib data. 

\subsection{Determination of the K-shell photoionization cross sections}
\begin{figure}[!h]
	\centering
	\includegraphics[width = 14cm]{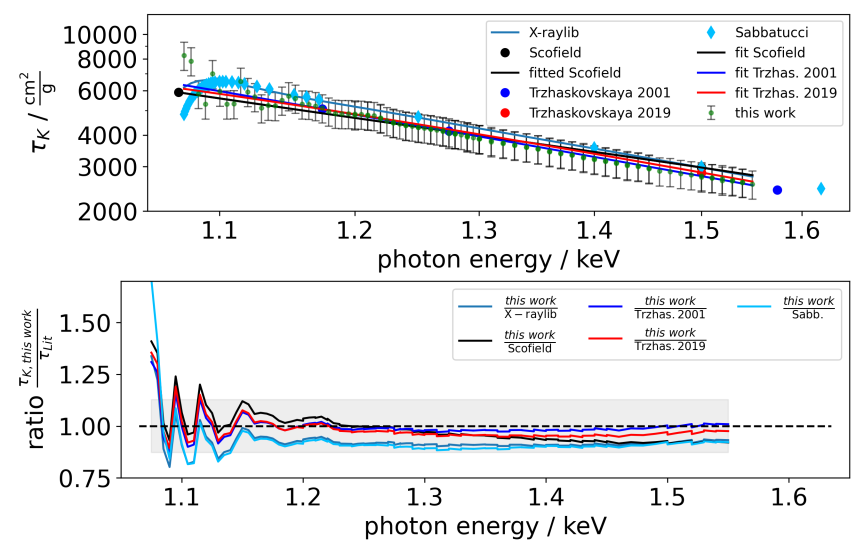}
    \caption{Comparison of the experimentally determined Na-K shell photoionization cross section in comparison to the X-raylib data \cite{T.Schoonjans2011} and different calculated cross sections \cite{Scofield1973, M.B.Trzhaskoskaya2001, Trzhaskovskaya_2019, L.Sabbatucci2015}. The lower part depicts the ratios between our experimental results and the respective calculation. Obviously, the fine structure related oscillations in our data are not taken into account in the calculations.}
	\label{fig:fig5}
\end{figure}

By considering the derived $\omega_{\text{F}}$, the photoionization cross sections for the sodium K-shell can be derived from the FPCSs shown above. Obviously, the same consequences of the areal mass determination affect the photoionization cross sections shown in Figure \ref{fig:fig5}. In addition to the determined cross sections, several available calculated photoionization cross sections for the sodium K-shell are shown as well \cite{Scofield1973, M.B.Trzhaskoskaya2001, Trzhaskovskaya_2019, L.Sabbatucci2015} as well as data from X-raylib \cite{T.Schoonjans2011}. For calculations, which only provide few data points in the shown photon energy range, the data was interpolated using Ebel polynomials for easier comparison to the experimental data. 

The x-ray absorption fine structure in the vicinity of the sodium K-edge is missing in the calculated datasets and the X-raylib data due to reasons discussed above. A more meaningful information that can be derived from these results is the photon energy dependence of the various cross section datasets. Slight deviations with respect to the experimentally derived data can be observed in the top panel of Figure \ref{fig:fig5}. For a more direct comparison, the bottom panel shows the ratios between our experimental results and the respective reference dataset. From this panel, the best agreement is found for the calculated data by Trzhaskovskaya \cite{M.B.Trzhaskoskaya2001}. Interestingly, the more recent data by the same authors \cite{Trzhaskovskaya_2019} shows slightly less agreement with respect to the absolute values. The agreement for the X-raylib \cite{T.Schoonjans2011} as well as the calculations by Sabbatucci \cite{L.Sabbatucci2015} is less in agreement with respect to the absolute values but of similar quality when considering the excitation photon energy dependence. The calculations by Scofield \cite{Scofield1973} are showing a different excitation energy dependence, which is in contrast to findings in our earlier works \cite{P.Hoenicke2014, H_nicke_2024}. But this may be an artifact due to only two data points being within this photon energy range.

It should be noted that there are other available databases for K-subshell photoionization cross sections, that employ the so-called edge jump ratio approach to separate the total photoionization cross section into individual subshell contributions. This approach results in constant subshell cross section ratios which are independent from the excitation photon energy and therefore, yielding wrong results for all subshells except K-shells. Thus we did not incorporate them into the comparison. 

\section{Conclusion}
In this work, we applied the holistic fundamental parameter determination approach towards the K-shell FPs of sodium using a thin NaCl coating on a SiN membrane. We derived the sodium K-shell fluorescence and Auger yields and validated different tabulated datasets for the sodium K-shell FPCSs and the K-shell photoionization cross sections for several hundred eV above the K-(1s) ionization threshhold.

Especially for the K-shell fluorescence yield, we were able to deduce a value with a reliable uncertainty budget of $0.0234 \pm 0.0015$, which is significantly lower compared to the Krause estimate of 10 \% \cite{Krause1979}. This enables a reduction of FP-related uncertainties for fundamental parameter based XRF quantification of sodium. In the case of sodium, we found that the X-raylib value is slightly underestimated. This is in contrast to our earlier findings, where X-raylib is usually the best agreeing database for X-ray interaction data. The determined FP data from this work can be downloaded from Zenodo for easier uptake \cite{Hoenicke2025Zen}.

\section*{Acknowledgments}
This work was supported by the European Partnership on Metrology, co-financed by the European Union's Horizon Europe Research and Innovation Programme and by the Participating States through grant agreement 21GRD01 (OpMetBat). In addition, the work was supported by the project 14ACMOS (grant agreement number 101096772), which is funded by the Chips Joint Undertaking and its
members, including the top-up funding of Belgium and the Netherlands. Parts of this research have been supported by Hi-Acts, an innovation platform under the grant of the Helmholtz Association HGF within the project BATIX.

\bibliographystyle{unsrt}
\bibliography{referencen}  %%% Uncomment this line and comment out the ``thebibliography'' section below to use the external .bib file (using bibtex) .

\begin{thebibliography}{10}

\bibitem{Yaroshevsky_2006}
A.~A. Yaroshevsky.
\newblock Abundances of chemical elements in the earth’s crust.
\newblock {\em Geochemistry International}, 44(1):48--55, January 2006.

\bibitem{Jaques_2023}
David~A. Jaques and Belen Ponte.
\newblock Dietary sodium and human health.
\newblock {\em Nutrients}, 15(17):3696, August 2023.

\bibitem{Shen_2023}
Yu-Fu Shen, S.~A. Alexeeva, Gang Zhao, Shuai Liu, Zeming Zhou, Hongliang Yan,
  Haining Li, Tianyi Chen, Xiaodong Xu, Huiling Chen, Huawei Zhang, and
  Jianrong Shi.
\newblock Sodium abundances in very metal-poor stars.
\newblock {\em Research in Astronomy and Astrophysics}, 23(7):075019, June
  2023.

\bibitem{M_ller_2011}
Margit~S. Müller.
\newblock A pinch of sodium.
\newblock {\em Nature Chemistry}, 3(12):974--974, November 2011.

\bibitem{Usiskin_2021}
Robert Usiskin, Yaxiang Lu, Jelena Popovic, Markas Law, Palani Balaya,
  Yong-Sheng Hu, and Joachim Maier.
\newblock Fundamentals, status and promise of sodium-based batteries.
\newblock {\em Nature Reviews Materials}, 6(11):1020--1035, June 2021.

\bibitem{Liu2023}
Zhaoguo Liu, Ziyang Lu, Shaohua Guo, Quan-Hong Yang, and Haoshen Zhou.
\newblock Toward high performance anodes for sodium-ion batteries: From hard
  carbons to anode-free systems.
\newblock {\em ACS Central Science}, 9(6):1076--1087, 2023.

\bibitem{WU2024}
Yujun Wu, Wei Shuang, Ya~Wang, Fuyou Chen, Shaobing Tang, Xing-Long Wu, Zhengyu
  Bai, Lin Yang, and Jiujun Zhang.
\newblock Recent progress in sodium-ion batteries: Advanced materials, reaction
  mechanisms and energy applications.
\newblock {\em Electrochemical Energy Reviews}, 7(17):2520--8136, 2024.

\bibitem{Krause1979}
M.O. Krause.
\newblock Atomic radiative and radiationless yields for k and l shells.
\newblock {\em J. Phys. Chem. Ref. Data}, 8(2):307--327, 1979.

\bibitem{FPI}
International initiative on x-ray fundamental parameters.
\newblock \url{https://www.exsa.hu/fpi.php}, 2021.
\newblock Accessed: 2022-06-03.

\bibitem{Menesguen2017}
Y.~M\'{e}nesguen, M.C. L\'{e}py, P.~H\"onicke, R.~Unterumsberger M.~M\"uller,
  B.~Beckhoff, J.~Hoszowska, and J.-Cl. Dousse.
\newblock Experimental determination of x-ray atomic fundamental parameters of
  nickel.
\newblock {\em Metrologia}, 2017.

\bibitem{Guerra2018}
M.~Guerra, J.~M. Sampaio, F.~Parente, P.~Indelicato, P.~Hönicke, M.~Müller,
  B.~Beckhoff, J.P. Marques, and J.P. Santos.
\newblock Theoretical and experimental determination of k- and l-shell x-ray
  relaxation parameters in ni.
\newblock {\em Phys. Rev. A}, 97:042501, 2018.

\bibitem{Baptista_2024}
Gonçalo Baptista, Daniel Pinheiro, Jorge Machado, Mauro Guerra, Pedro Amaro,
  and José~Paulo Santos.
\newblock Lisbon atomic database (lisa): a compilation of calculated
  fundamental atomic parameters.
\newblock {\em The European Physical Journal D}, 78(2), February 2024.

\bibitem{Beckhoff_2022}
B.~Beckhoff.
\newblock Traceable characterization of nanomaterials by x-ray spectrometry
  using calibrated instrumentation.
\newblock {\em Nanomaterials}, 12(13):2255, jun 2022.

\bibitem{P.Hoenicke2016}
P.~Hönicke, M.~Kolbe, and B.~Beckhoff.
\newblock What are the correct l-subshell photoionization cross sections for
  quantitative x-ray spectroscopy?
\newblock {\em X-Ray Spectrom.}, 45(4):207--211, 2016.

\bibitem{P.Hoenicke2016a}
P.~H\"onicke, M.~Kolbe, M.~Krumrey, R.~Unterumsberger, and B.~Beckhoff.
\newblock Experimental determination of the oxygen k-shell fluorescence yield
  using thin sio2 and al2o3 foils.
\newblock {\em Spectrochim. Acta B}, 124:94--98, 2016.

\bibitem{Kayser_2022}
Y.~Kayser, P.~Hönicke, M.~Wansleben, A.~Wählisch, and B.~Beckhoff.
\newblock Experimental determination of the gadolinium l subshells fluorescence
  yields and coster-kronig transition probabilities.
\newblock {\em X-Ray Spectrom.}, nov 2022.

\bibitem{H_nicke_2023}
Philipp Hönicke.
\newblock A novel and holistic approach for experimental x-ray fundamental
  parameter determination - the ru l-shell.
\newblock {\em New Journal of Physics}, 25(7):073012, jul 2023.

\bibitem{H_nicke_2024}
Philipp Hönicke.
\newblock Experimental determination of the sulfur k-shell fundamental
  parameters employing the holistic approach.
\newblock {\em New J. Phys.}, 26(10):103011, September 2024.

\bibitem{Wauschkuhn_2024}
Nils Wauschkuhn, Heiko Gundlach, and Philipp Hönicke.
\newblock Experimental determination of the hafnium l-subshell fundamental
  parameters using the holistic approach.
\newblock {\em New J. Phys.}, 26(3):033017, March 2024.

\bibitem{Hoenicke2025Zen}
Philipp Hönicke.
\newblock Experimental determination of the sodium k-shell atomic fundamental
  parameters for x-ray spectroscopy, 2025.

\bibitem{J.Lubeck2013}
J.~Lubeck, B.~Beckhoff, R.~Fliegauf, I.~Holfelder, P.~H\"onicke, M.~M\"uller,
  B.~Pollakowski, F.~Reinhardt, and J.~Weser.
\newblock A novel instrument for quantitative nanoanalytics involving
  complementary x-ray methodologies.
\newblock {\em Rev. Sci. Instrum.}, 84:045106, 2013.

\bibitem{F.Senf1998}
F.~Senf, U.~Flechsig, F.~Eggenstein, W.~Gudat, R.~Klein, H.~Rabus, and G.~Ulm.
\newblock A plane-grating monochromator beamline for the ptb undulators at
  {BESSY} {II}.
\newblock {\em J. Synchrotron Rad.}, 5:780--782, 1998.

\bibitem{F.Scholze2009}
F.~Scholze and M.~Procop.
\newblock Modelling the response function of energy dispersive x-ray
  spectrometers with silicon detectors.
\newblock {\em X-Ray Spectrom.}, 38(4):312--321, 2009.

\bibitem{H.Ebel2003}
H.~Ebel, R.~Svagera, M.F. Ebel, A.~Shaltout, and J.H. Hubbell.
\newblock Numerical description of photoelectric absorption coefficients for
  fundamental parameter programs.
\newblock {\em X-Ray Spectrom.}, 32:442--451, 2003.

\bibitem{Sherman1955}
J.~Sherman.
\newblock The theoretical derivation of fluorescent x-ray intensities from
  mixtures.
\newblock {\em Spectrochim. Acta}, 7:283--306, 1955.

\bibitem{Beckhoff2008}
B.~Beckhoff.
\newblock Reference-free x-ray spectrometry based on metrology using
  synchrotron radiation.
\newblock {\em J. Anal. At. Spectrom.}, 23:845 -- 853, 2008.

\bibitem{HfPaper}
Nils Wauschkuhn, Heiko Gundlach, and Philipp Hönicke.
\newblock Experimental determination of the hafnium l-subshell fundamental
  parameters using the holistic approach.
\newblock {\em New J. Phys.}, 26(3):033017, March 2024.

\bibitem{Foreman_Mackey_2013}
D.~Foreman-Mackey, D.W. Hogg, D.~Lang, and J.~Goodman.
\newblock emcee: The {MCMC} hammer.
\newblock {\em Publications of the Astronomical Society of the Pacific},
  125(925):306--312, mar 2013.

\bibitem{T.Schoonjans2011}
T.~Schoonjans, A.~Brunetti, B.~Golosio, M.~Sanchez del Rio, V.A. Sol\'{e},
  C.~Ferrero, and L.~Vincze.
\newblock The xraylib library for x-ray–matter interactions. recent
  developments.
\newblock {\em Spectrochim. Acta B}, 66:776 -- 784, 2011.

\bibitem{Bambynek1984}
W.~Bambynek.
\newblock New evaluation of k-fluorescence yields.
\newblock {\em X-Ray and Inner-Shell Processes in Atoms, Molecules and
  Solids,Post-Deadline Abstracts}, page~1, 1984.

\bibitem{W.T.Elam2002}
W.T. Elam, B.D. Ravel, and J.R. Sieber.
\newblock A new atomic database for x-ray spectroscopic calculations.
\newblock {\em Rad. Phys. Chem.}, 63:121--128, 2002.

\bibitem{A.Rani1988}
A.~Rani, R.K. Koshal, S.N. Chaturvedi, and N.~Nath.
\newblock Photon-excited k x-ray fluorescence cross-section measurements for
  some low-z-elements.
\newblock {\em X-Ray Spectrom.}, 17(2):53--54, 1988.

\bibitem{S.Daoudi2015}
S.~Daoudi, A.~Kahoul, Y.~Sahnoune, B.~Deghfel, Y~Kasri, F.~Khalfallah,
  V.~Aylikci, N.K. Aylikci, D.E. Medjadi, and M.~Nekkab.
\newblock New k-shell fluorescence yields curve for elements with
  3$\leq$z$\leq$99.
\newblock {\em Journal of the Korean Physical Society}, 67(9):1537--1543, 2015.

\bibitem{J.H.Hubbell1994}
J.H. Hubbell, P.N. Trehan, N.~Singh, B.~Chand, D.~Mehta, M.L. Garg, R.R. Garg,
  S.~Singh, and S.~Puri.
\newblock A review, bibliography and tabulation of k, l and higher atomic shell
  x-ray fluorescence yields.
\newblock {\em J. Phys. Chem. Ref. Data}, 23(2):339--364, 1994.

\bibitem{S.T.Perkins1991}
S.T. Perkins, D.E. Cullen, M.H. Chen, J.~Rathkopf, J.~Scofield, and J.H.
  Hubbell.
\newblock Tables and graphs of atomic subshell and relaxation data derived from
  the llnl evaluated atomic data library (eadl), z = 1-100.
\newblock {\em UCRL 50400}, 30, 1991.

\bibitem{V.O.Kostroun1971}
V.O. Kostroun, M.H. Chen, and B.~Crasemann.
\newblock Atomic radiation transition probabilities to the 1s state and
  theoretical k-shell fluorescence yields.
\newblock {\em Phys. Rev. A}, 3(2):533--545, 1971.

\bibitem{Walters_1971}
D.~L. Walters and C.~P. Bhalla.
\newblock Nonrelativistic auger rates, x-ray rates, and fluorescence yields for
  the k-shell.
\newblock {\em Physical Review A}, 3(6):1919--1927, jun 1971.

\bibitem{McGuire_1969}
Eugene~J. McGuire.
\newblock K-shell auger transition rates and fluorescence yields for elements
  be-ar.
\newblock {\em Physical Review}, 185(1):1--6, sep 1969.

\bibitem{Zschornack}
G.H. Zschornack, editor.
\newblock {\em Handbook of X-ray Data}.
\newblock Springer, {N}ew {Y}ork, 2007.

\bibitem{Zschornack2007}
G.~Zschornack.
\newblock {\em Handbook of X-ray data}.
\newblock Springer-Verlag, 2007.

\bibitem{Scofield1973}
J.H. Scofield.
\newblock Theoretical photoionisation cross sections from 1 to 1500 kev.
\newblock {\em UCRL-51326}, 1973.

\bibitem{M.B.Trzhaskoskaya2001}
M.B. Trzhaskovskaya, V.I. Nefedov, and V.G. Yarzhemsky.
\newblock Photoelectron angular distribution parameters for elements z=1 to
  z=54 in the photo electron energy range 100 ev to 5000 ev.
\newblock {\em Atom. Data Nucl. Data}, 77:97--159, 2001.

\bibitem{Trzhaskovskaya_2019}
M.B. Trzhaskovskaya and V.G. Yarzhemsky.
\newblock Dirac-fock photoionization parameters for {HAXPES} applications, part
  {II}: Inner atomic shells.
\newblock {\em Atomic Data and Nuclear Data Tables}, 129-130:101280, sep 2019.

\bibitem{L.Sabbatucci2015}
L.~Sabbatucci and F.~Salvat.
\newblock Theory and calculation of the atomic photoeffect.
\newblock {\em Radiat. Phys. Chem.}, 121:122--140, 2015.

\bibitem{P.Hoenicke2014}
P.~H\"onicke, M.~Kolbe, M.~M\"uller, M.~Mantler, M.~Kr\"amer, and B.~Beckhoff.
\newblock Experimental verification of the individual energy dependencies of
  the partial l-shell photoionization cross sections of pd and mo.
\newblock {\em Phys. Rev. Lett.}, 113(16):163001, 2014.

\end{thebibliography}





% \end{thebibliography}

\end{document}